# Is the Time-Delay Signal in Bose-Einstein Correlations a Signature for the Quark-Gluon Plasma at RHIC ?

B.R. Schlei[1]

[1] Theoretical Division, T-1, Los Alamos National Laboratory, Los Alamos, USA

In 1996, D.H. Rischke and M. Gyulassy proposed to use the time-delay signal in Bose-Einstein correlations (BEC) as a signature for the possible formation of a quark-gluon plasma (QGP) in relativistic heavy-ion collisions. In particular, they suggested to measure the ratio of the transversal interferometry radii, $R_{out}/R_{side}$, which can be obtained by fitting experimental BEC data with a parametrization introduced by G. Bertsch et al. in 1988. The transverse radius parameter $R_{out}$ has compared to the transverse radius parameter $R_{side}$ an additional temporal dependence which should be sensitive to a prolonged lifetime of a fireball, in case a QGP was formed in a relativistic heavy-ion collision. In my presentation, I am going to revisit the above made considerations by addressing various possible scenarios for hadron spectra (single and double inclusives) for Pb+Pb collisions at CERN/SPS beam energies and for Au+Au collisions at BNL/RHIC beam energies, respectively, within a relativistic *true* hydrodynamic framework (i.e., HYLANDER-C). I claim, that the time-delay signal in Bose-Einstein correlations *is not* a signature for the possible formation of a QGP, and I am going to explain why this is the case.

# Contents:

1. **QGP Stall**; and some further remarks on **Bose-Einstein Correlations** (BEC).

2. relativistic **Hydrodynamics** and the **Equation Of State** (EOS) of nuclear matter.

3. from **CERN / SPS** to **BNL / RHIC**: examples for **E dN/dp** and **BEC** spectra.

4. my **Predictions** for RHIC regarding the **QGP Stall**.

# HBT - A Signature for QGP ?

The ratio of the transverse interferometry radii R_out / R_side is a function of the excitation energy density, due to D.H. Rischke, et al.

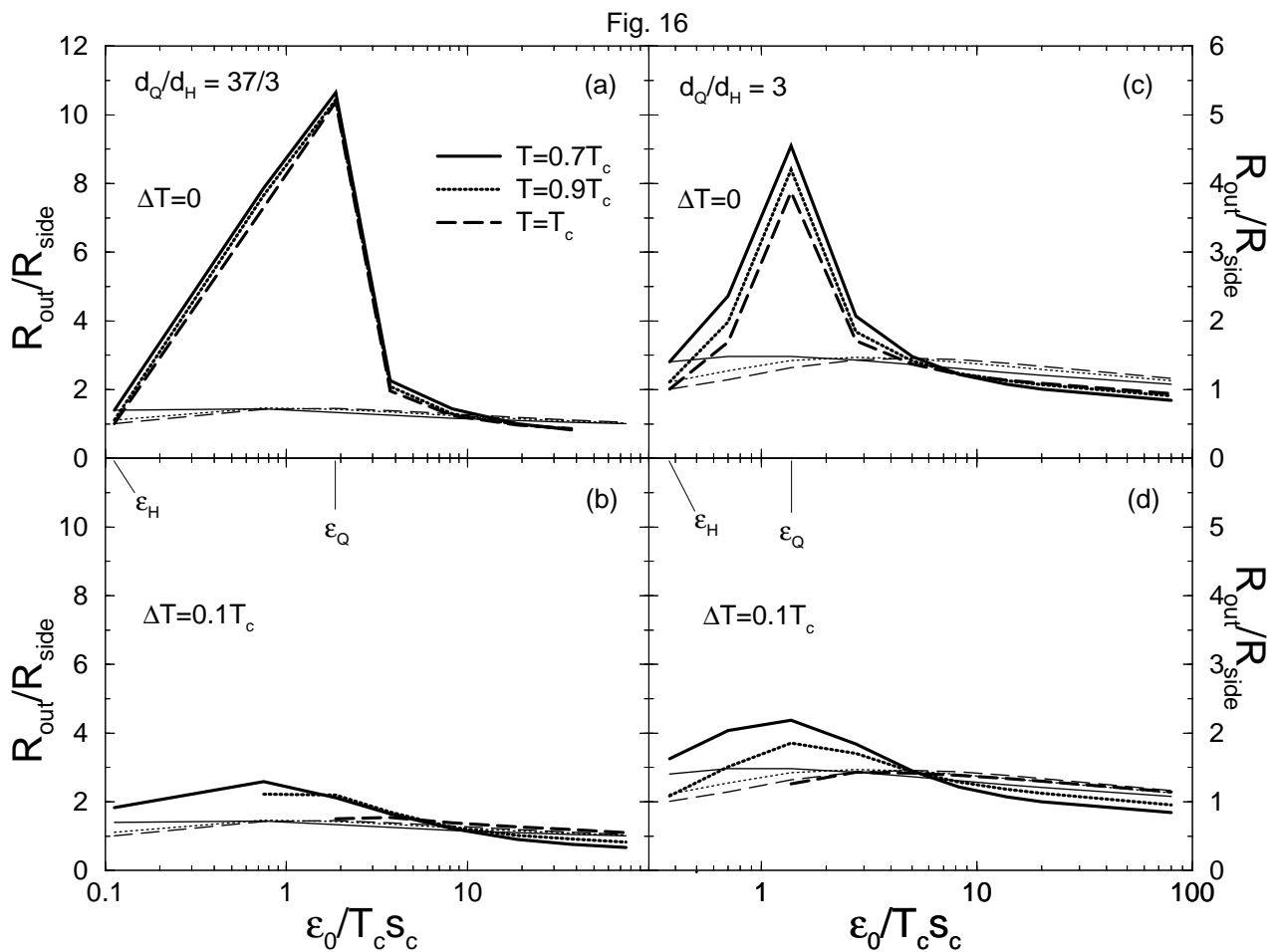

Fig. 16

**Figure taken from:**

D.H. Rischke, M. Gyulassy, Nucl. Phys. A608 (1996) 479.

# Hydrodynamics and EOS

**1.** fully-fledged **Hydrodynamics** is **constrained**.

**2.** effective **Softness** determines $m_\perp$ spectra **slopes**.

B.R. Schlei, D. Strottman, N. Xu, Phys. Rev. Lett. 80 (1998) 3467.

**3.** **BEC** is mostly determined by $T_f(\varepsilon_f)$.

B.R. Schlei, CF'98 Proceedings; B.R. Schlei et al., LA-UR-98-4184.

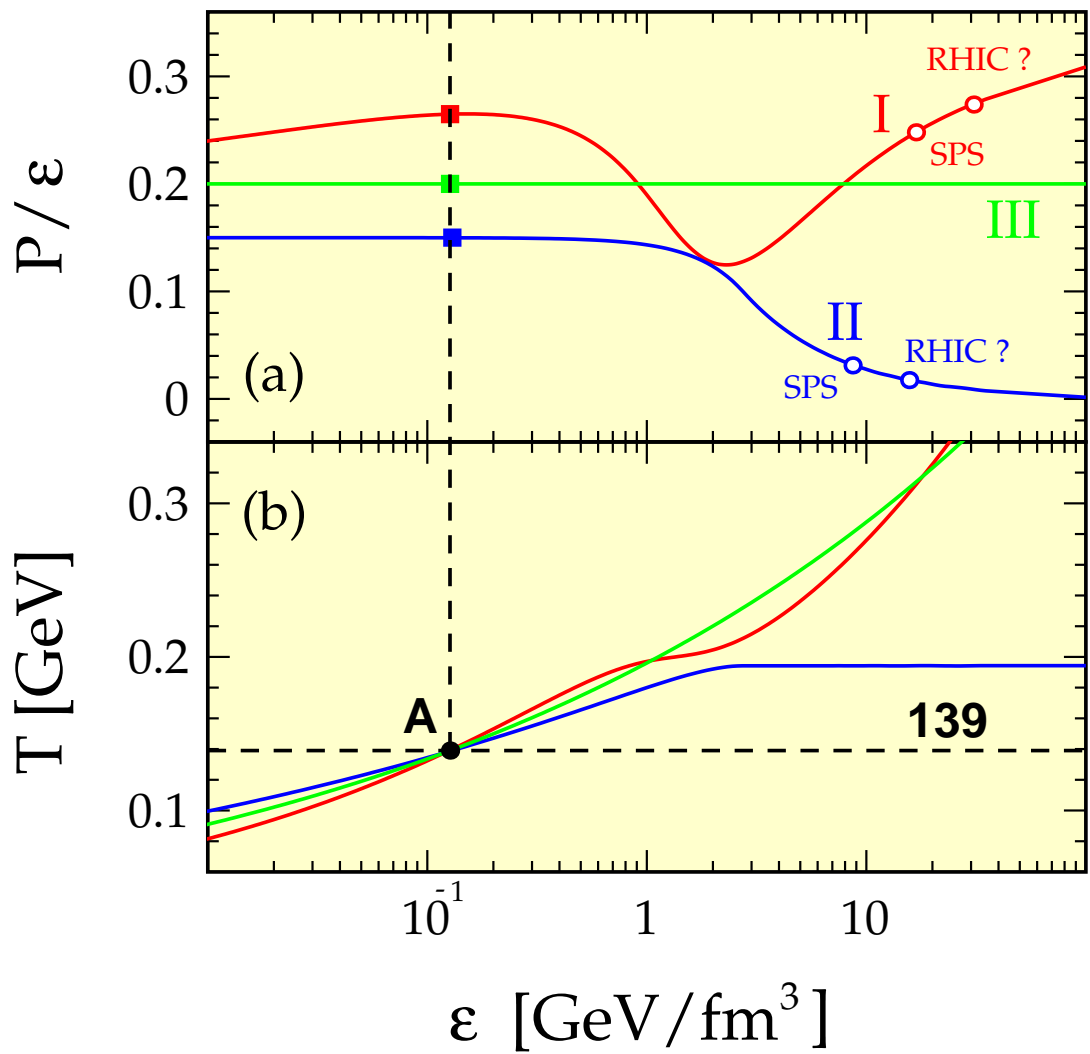

# Earlier Results: CERN / SPS

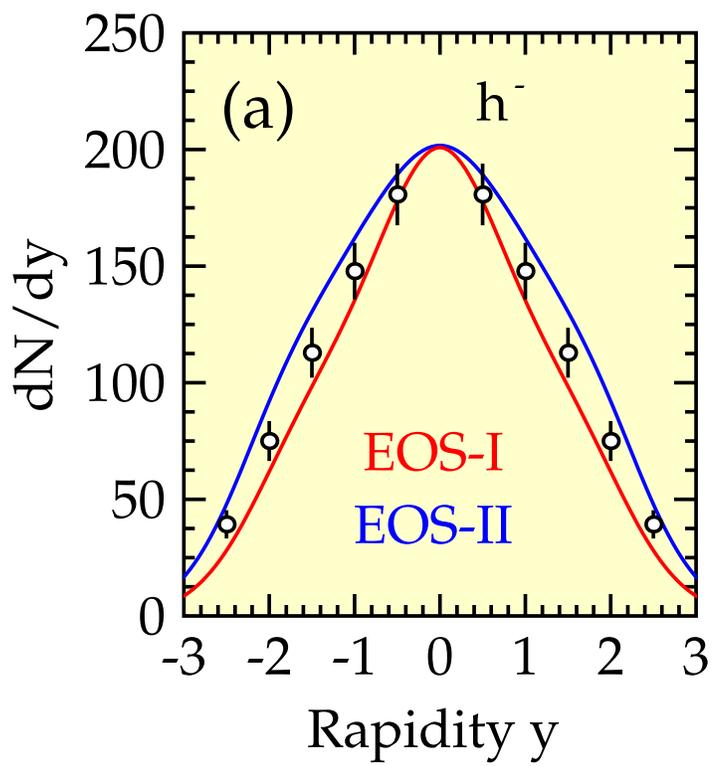
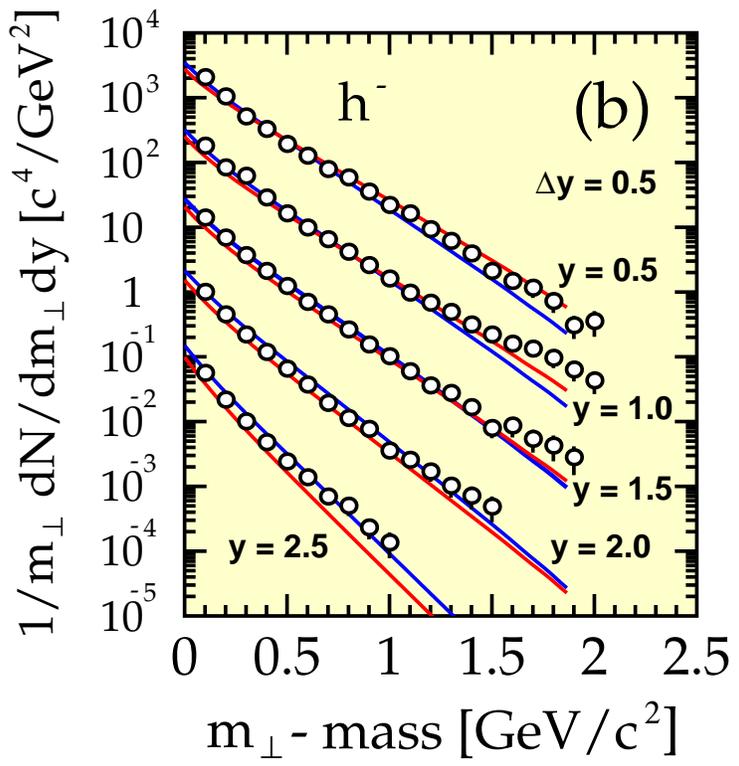
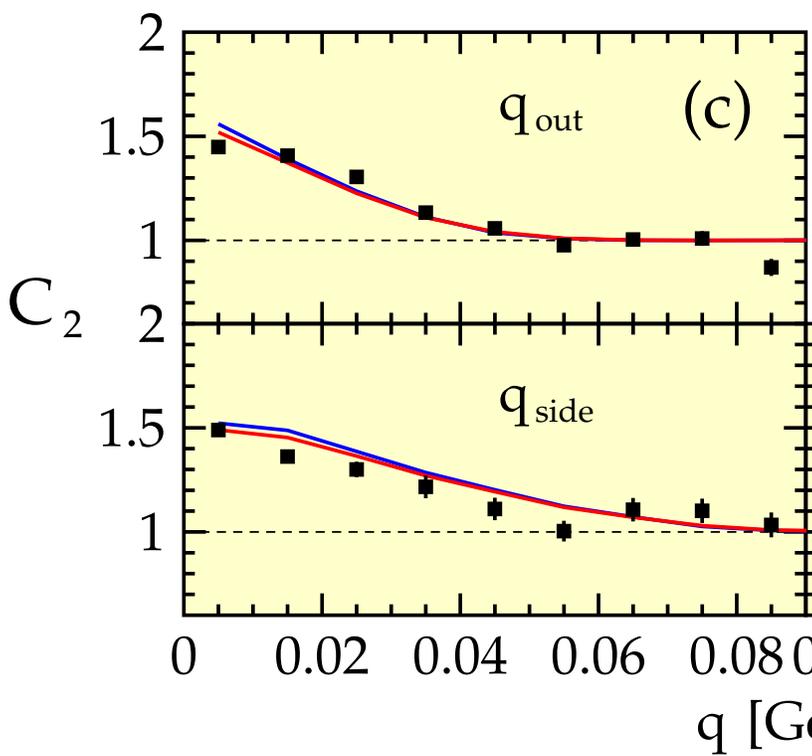

**Singles:** NA49 preliminary
**BEC:** NA44

**R_out / R_side ~ 1.15**

# HYLANDER-C @ "RHIC ?"

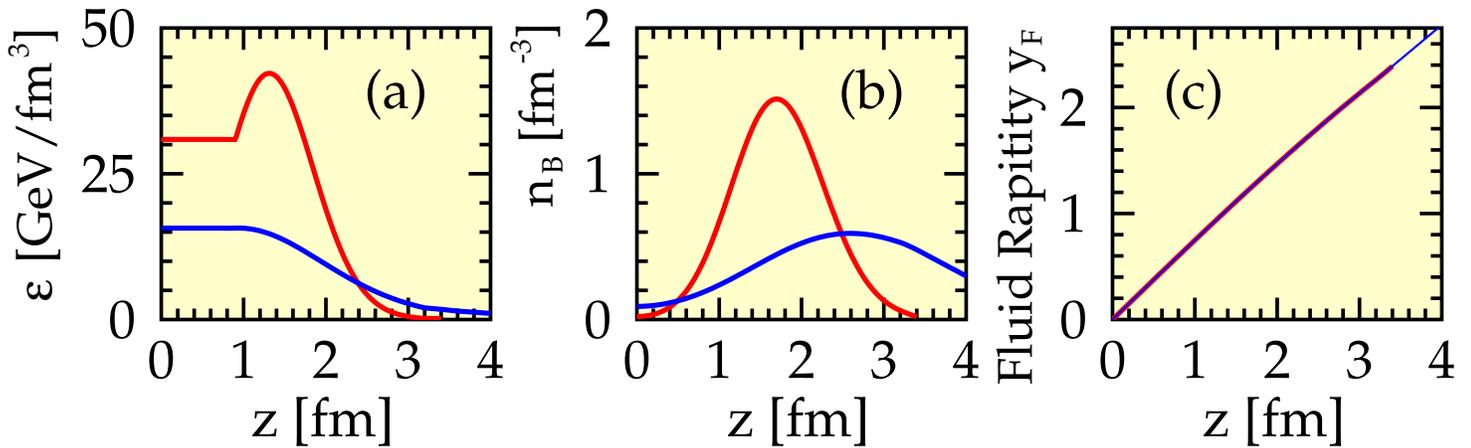

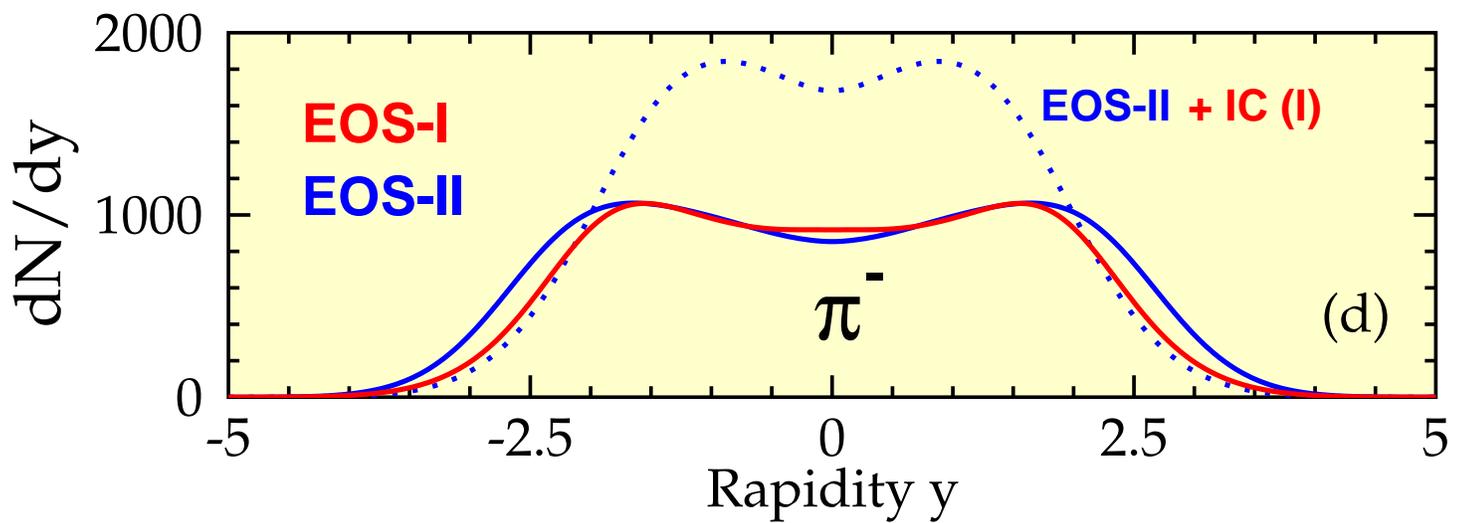

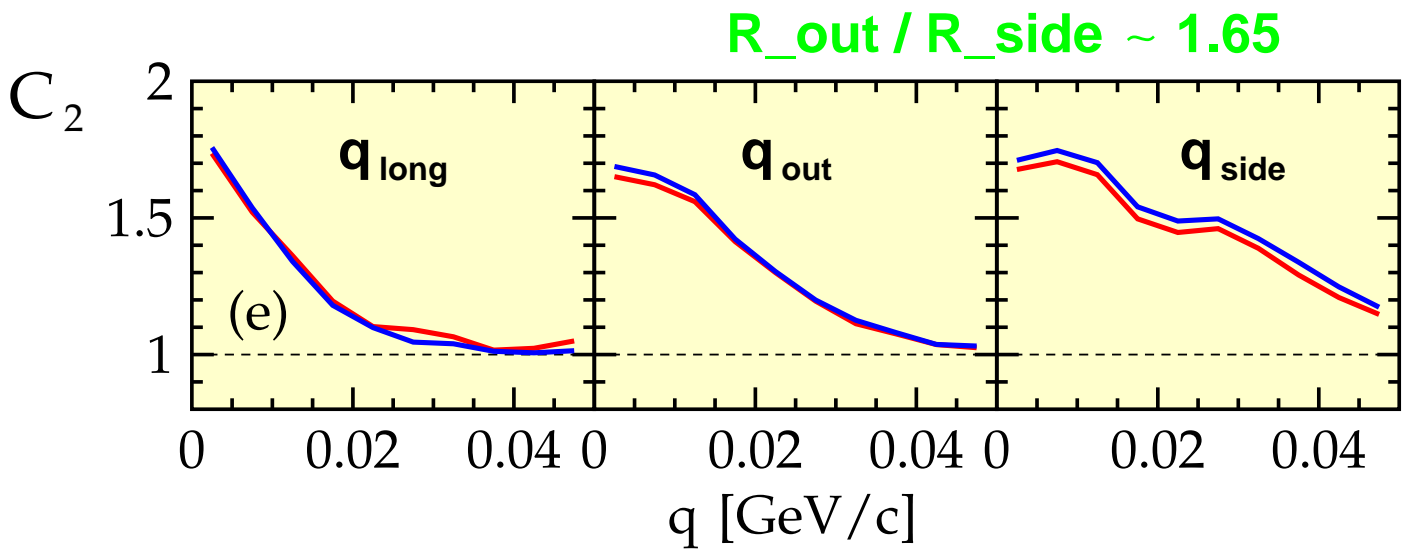

# Analysis and Conclusions

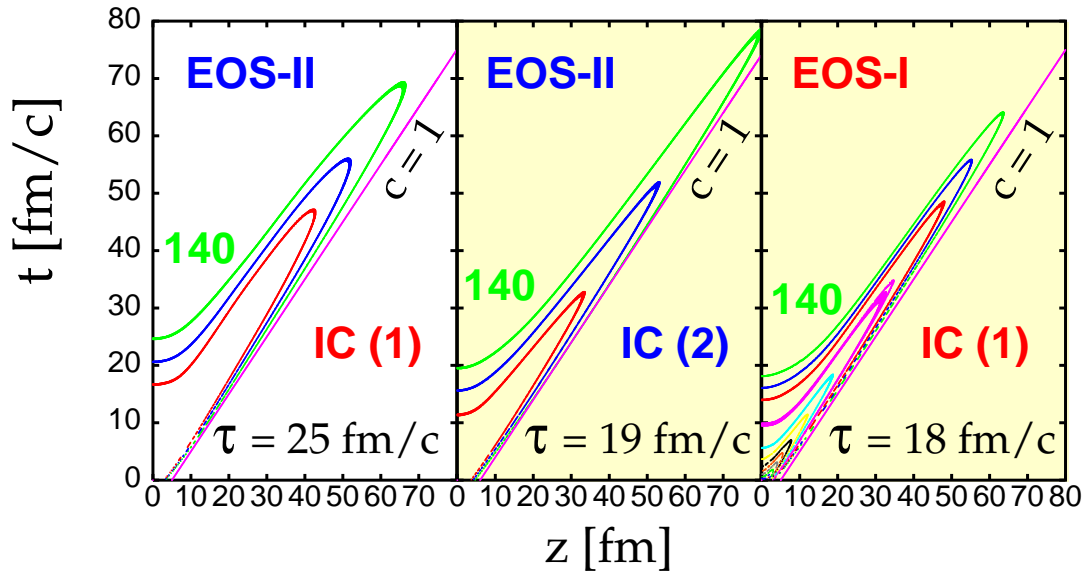

1. similar **E dN/dp** leads to similar **space-time geometry** through adjustment of **initial conditions**.

2. **R_out / R_side** changes with **initially** achieved $\varepsilon_o$.

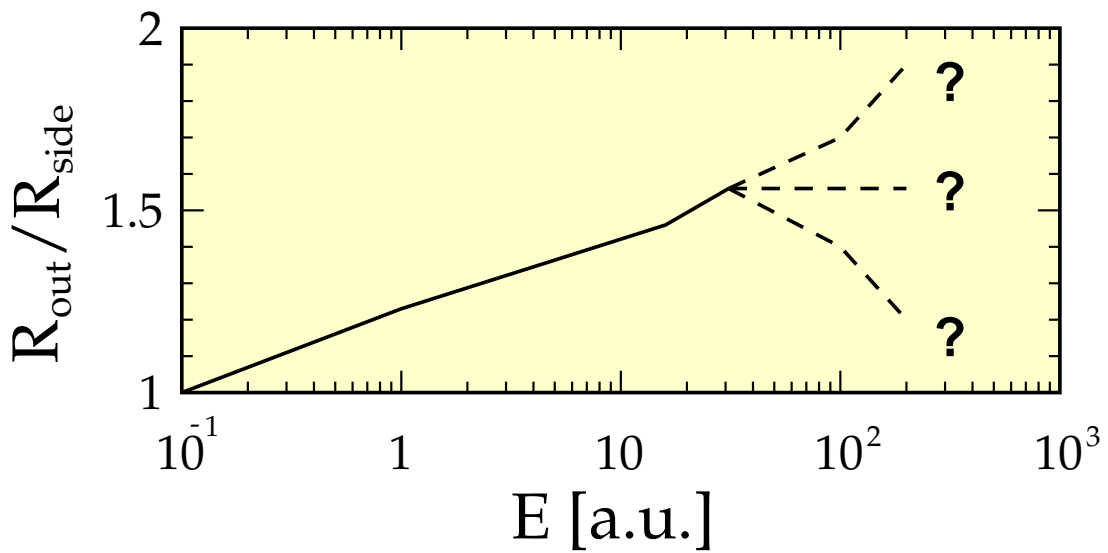

**!!! BEC are not a QGP signature !!!**